\documentclass[prb,twocolumn,preprintnumbers,amsmath,amssymb,superscriptaddress]{revtex4}

\usepackage{graphicx}
\usepackage{dcolumn}
\usepackage{bm}

\begin{document}

\title{Few-layer graphene on SiC, pyrolitic graphite, and graphene: A Raman scattering study}
\author{C. \surname{Faugeras}$^{}$}
\email{clement.faugeras@grenoble.cnrs.fr}
\author{A. \surname{Nerri\`ere}$^{}$}
\author{M. \surname{Potemski}$^{}$}
\affiliation{Grenoble High Magnetic Field Laboratory, CNRS, 38042
Grenoble, France }

\author{A  \surname{Mahmood}$^{}$}
\author{E. \surname{Dujardin}$^{}$}
\affiliation{CEMES, CNRS, 31055 Toulouse, France }

\author{C. \surname{Berger }$^{}$}
\author{W. A. \surname{de Heer }$^{}$}
\affiliation{Georgia Institute of Technology, Atlanta, Georgia,
USA }


\begin{abstract}
The results of micro-Raman scattering measurements performed on
three different ``graphitic'' materials: micro-structured disks of
highly oriented pyrolytic graphite, graphene multi-layers
thermally decomposed from carbon terminated surface of 4H-SiC and
an exfoliated graphene monolayer are presented. Despite its
multi-layer character, most parts of the surface of the
graphitized SiC substrates shows a single-component, Lorentzian
shape, double resonance Raman feature in striking similarity to
the case of a single graphene monolayer. Our observation suggests
a very weak electronic coupling between graphitic layers on the
SiC surface, which therefore can be considered to be graphene
multi-layers with a simple (Dirac-like) band structure.
\end{abstract}

\maketitle

The interest in the properties of graphite-like allotropes of
carbon has recently been rekindled when its simplest form,
graphene, a mono layer of carbon atoms arranged in a honey comb
lattice, was experimentally identified and readily produced by a
simple exfoliation technique \cite{Novoselov04}. The linear
dispersion at the K and K' points of the graphene band structure
gives rise to a relativistic-like behavior of free carriers that
has many implications for the observed quantum mechanical effects,
such as the half-integer quantum Hall effect, and has stimulated
considerable research over the past few years
\cite{Novoselov05,Zhang05a,Geim07}. Alternatively, thin graphitic
layers can be epitaxially grown on 4H-SiC substrate
\cite{Berger04,Berger06}, by the thermal decomposition of either
Si- or C-terminated surface. In this case, the material, herein
referred to as few-layer graphene (FLG) on SiC, may be composed of
one or many graphene layers, depending on the growth conditions.
Although much interesting physics has emerged from the studies of
exfoliated graphene structures, the epitaxy of graphene seems to
be the most viable production route for electronic applications
\cite{Geim07}.

In this letter, we present the results of micro-Raman scattering
studies of few-layer graphene on C-terminated surface of SiC
(FLG-C-SiC) with a large number of layers, and compare them with
the data obtained on disks made of highly oriented pyrolitic
graphite (HOPG) with different thicknesses and with the spectra of
an exfoliated graphene monolayer. Attention is focused on the form
of the so called D'-band, a well established feature in the Raman
scattering spectra of graphite\cite{Reich04} and
graphene\cite{Ferrari06,Graf07,Gupta06}. Its single-component form
is the fingerprint of the simple electronic bands in a graphene
monolayer, while a multi-component form is the signature of the
complex band structure of multi-layers with Bernal-type
(graphite-like) stacking.

Apart from some residual inclusions, most of the surface of the
FLG-C-SiC probed with micro-Raman spectroscopy was found to show a
single component D' feature, as for an exfoliated graphene
monolayer but in striking contrast to the double-component
D'-feature observed in HOPG disks. This observation supports the
view that FLG-C-SiC is a system which displays single band
(Dirac-like) dispersion relation of electronic states
characteristic of decoupled graphene layers.

The structures used for experiments were disks of HOPG with
different thickness, an exfoliated graphene monolayer, and, three
FLG-C-SiC samples with a different number of layers. HOPG disks
(20-$\mu$m in diameter) have been prepared by mechanical cleavage
with a pulled glass tip of cylindrical ZYA grade HOPG mesas and by
deposition via micro-manipulation onto a silicon substrate with a
100-nm or 300-nm SiO$_2$ layer \cite{Zhang05b,Mahmood07}.
Graphitic disks could be further thinned by exfoliation down to a
thickness ranging between 2 and 20 nm. A graphene monolayer was
prepared by exfoliation of freshly cleaved ZYA grade HOPG on a
silicon substrate with a 300 nm thermal oxide substrate as
described in the literature \cite{Novoselov04}. Sub-20 nm thick
graphitic disks and the graphene layer were subsequently
characterised/identified by AFM. FLG-C-SiC structures were
thermally decomposed from 4H-SiC substrates following the
procedure detailed in Ref. 5 and 6. Two of them (sample 1 and 2)
are 5-10 layers thick while sample 3 consists of 70-90 layers, as
determined from the intensity ratio of the Si 92-eV and C 271-eV
Auger peaks \cite{Berger04}. Raman scattering spectra have been
measured using a confocal microscope with 2$\mu$m spatial and 1 nm
spectral resolution. Experiments were performed at room
temperature using the 632.8 nm line of a HeNe laser.

We start the discussion of the experimental results by presenting
the characteristic Raman-scattering spectrum of bulk graphite,
obtained here for a thick ($\sim$100 nm-high) disk of HOPG (see
Fig.~1). The G band at 1582 cm$^{-1}$ is due to first order (one
phonon) Raman scattering process. It is characteristic of sp$^{2}$
hybridization and involves the in-plane optical phonon E$_{2g}$
near the $\Gamma$ point of the phonon band structure. As a rule,
the overtone (two-phonon process) of the G-band, labeled as the
G'-band in Fig.~1 is much less intense. The first order Raman
scattering process due to characteristic zone boundary phonons (at
1325 cm$^{-1}$) is forbidden in defect-free structures and is
barely visible in our spectra. However, these phonons effectively
contribute to a double resonance Raman scattering
process\cite{Reich04} (involving two phonons) which results in the
appearance of the characteristic D' band at $\sim$2650 cm$^{-1}$
(under HeNe laser excitation). Such a Raman scattering process,
schematically shown in the inset of Fig.~1, is inherently
sensitive to the multi or single-band character of the dispersion
relations of the electronic states. In consequence, the D'-feature
is known to be composed of at least two-components in graphite (as
seen in Fig.~1) and also in other graphene multi-layers with
Bernal stacking due to the multi-band character of the electronic
states in these systems. In contrast, the unique single Lorentzian
form of the D'-feature is the signature of a system with a
single-band electronic dispersions, such as the representative
graphene monolayer.

\begin{figure}[]
\includegraphics[width=0.7\linewidth,angle=0,clip]{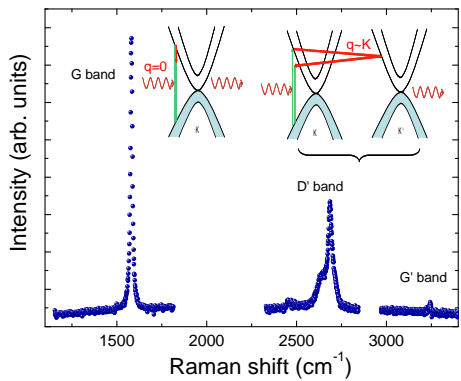}
\caption{Raman spectrum of a 20-$\mu$m diameter HOPG disk. Inset:
Schematics of the first and second order Raman scattering
processes responsible for the G band and the D' and G' bands
respectively. The phonon wave vector implied in both scattering
processes is indicated in red.} \label{fig1}
\end{figure}

Fig.~2 illustrates the form of the D'-band in Raman scattering
spectra measured for our HOPG disks with different thicknesses. A
detailed analysis of the observed changes of the form of the
D'-band is beyond the scope of this work, nevertheless it is clear
that the low energy part of this band gains in intensity when the
disk height is reduced down to 2 nm ($\sim$6 graphitic planes). To
quantify this change, we assume that this band is a simple
superposition of two Lorentzian peaks. The shape of the observed
D' band can be reasonable reproduced assuming the high energy peak
is fixed at 2686 cm$^{-1}$ but the center of the low energy peak
shifts from 2636 cm$^{-1}$ for HOPG to 2653 cm$^{-1}$ for a 2-nm
thick disk. The ratio of the amplitude of these two peaks can be
used to estimate the thickness of the graphite disks (see inset of
the Fig.~2). The present experiments performed on graphite with
thickness between 2 and 20 nm bridge the current measurement gap
between the well-documented bulk graphite \cite{Reich04} and
recent micro Raman scattering studies on exfoliated few layer
graphene\cite{Ferrari06,Graf07,Gupta06}.\\

\begin{figure}[]
\includegraphics[width=0.7\linewidth,angle=0,clip]{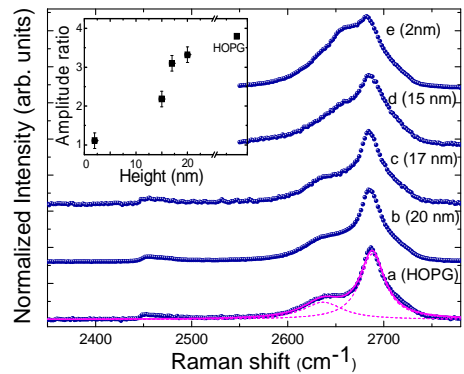}
\caption{Raman spectra of the D' feature of (a) HOPG and of 20
$\mu$m diameter HOPG thin disks with heights measured by AFM of
(b) 20 nm, (c) 17 nm, (d) 15 nm, (e) 2 nm. The solid and dashed
lines are Lorentzian fits. These spectra have been normalized to
the high energy component of the D' line to show the observed
effect. Inset : Amplitude ratio of the two components of the D'
line as a function of the height of the graphite disk.}
\label{fig2}
\end{figure}

We turn our attention now to the main topic of this paper which is
the investigations of FLG-C-SiC structures. Our results of micro
Raman scattering studies are summarized in Fig.~3. As a rule, the
Raman scattering spectra of these samples show the characteristic
G and D' bands on the background of a more or less pronounced
signal from the SiC substrate\cite{Leburton99}, and the spectrum
of the latter is shown for comparison (trace (a) in Fig.~3). When
scanning the laser spot over the FLG-C-SiC surface we find that
most of the sample gives rise to the characteristic Raman
scattering spectra which are shown in Fig.~3 (traces b, c and d),
for three different samples. The important feature of these
spectra is the appearance of the D' band in the form of a single
component peak. We note, however, the existence of some inclusions
or spots on the sample surface, which are visible under an optical
microscope, and which give rise to a different spectrum
characterized by a distinct, more complex shape of the D'-band
(see trace (b') in Fig.~3).

The Raman scattering spectra of our FLG-C-SiC samples can be
compared to the characteristic spectrum of the graphene monolayer
(trace (e) in Fig.~3) and/or to the spectra of HOPG disks shown in
Fig.~2. Within experimental uncertainty, the energetic position of
the G-peak is common for all samples. Not surprisingly, this means
that $E_{2g}$ in-plane optical phonons are the same in all these
materials. It is worth noticing that the spectra presented here
show no sign of the D-band. This demonstrate the weak disorder in
our samples\cite{Cancado06,Pimenta07}. The D-band can also appear
in the spectra measured at the edges of the graphitic
flakes\cite{Ferrari06}, but in our experiments the laser spot was
always focused on the interior of the measured structures.

\begin{figure}[]
\includegraphics[width=0.5\linewidth,angle=0,clip]{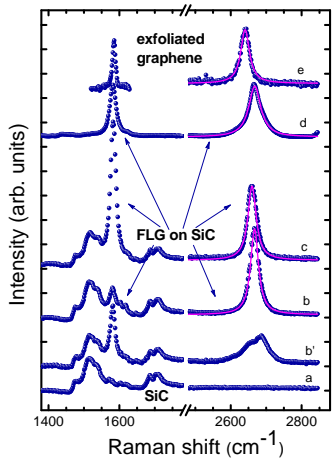}
\caption{Raman spectra of a 4H-SiC substrate (a), of a graphitic
residue on sample 1 (b'), of 5-10 layers FLG on 4H-SiC substrate
(b) for sample 1 and (c) for sample 2, of 70-90 layers epitaxial
FLG (d) and of exfoliated graphene (e). The solid lines are
Lorentzian fits.} \label{fig3}
\end{figure}

The resemblance of the shape of the D'-band in the Raman
scattering signal arising from the majority of the surface of our
FLG-C-SiC samples to the D'-band measured on the exfoliated
graphene flake is remarkable. A single Lorentzian peak is seen for
both type of structures. This is particularly clear for the
graphene and the FLG-C-SiC samples 1 and 2, each showing the same
half-width of 29 cm$^{-1}$. The fit of the Lorentzian profile to
the D'-line is not as good for the case of sample 3 which has as
many $\sim$90 layers, and results in a larger half-width (D = 40
cm$^{-1}$). The FLG-C-SiC samples differ however from graphene as
far as the position of the D'-band is concerned. In our spectra
for exfoliated graphene, this band is centered at 2641 cm$^{-1}$,
whereas the D'-peak of FLG-C-SiC samples appears at slightly
higher energies. Its actual position fluctuates from one location
to another of the laser spot on the sample and from sample to
sample, but remains in the range 2655 - 2665 cm$^{-1}$. The Raman
scattering signal which arises from the visible inclusions on
FLG-C-SiC surfaces shows a D'-band of a more complex form. We note
the resemblance to the D'-band characteristic of graphite in
general, and more precisely to the one observed for our thin HOPG
disks. We speculate that this minority Raman scattering signal
(linked to the inclusions) originates from some graphitic residues
(Bernal-stacked graphene multilayers) on the surface of our
FLG-C-SiC. Decomposing the D'-band characteristic of the graphitic
residue in samples 1 into two Lorentzian peaks we find that the
amplitude ratio of these peaks is the same as in the case of a
HOPG disk with a height of $\sim$3 nm ($\sim$10 monolayers).

The striking correspondence in the shape of the D'-band observed
in the majority Raman scattering signal from the FLG-C-SiC samples
and from graphene is a clear indication of the similar electronic
structures in both these material systems. This strongly supports
previous work pointing towards the quasi-two dimensional
Dirac-like character of electronic states in FLG-C-SiC samples
\cite{Berger06,Sadowski06}. Our results are also consistent with
recent reports which illustrate that two dimensional carbon layers
on C-face SiC substrates are not Bernal stacked. Instead, these
layers are generally rotated at specific angles \cite{Hass07}, and
in consequence are electronically well decoupled one with respect
to another. To some extent, the FLG-C-SiC resembles the so-called
``turbostratic'' graphite, also composed of non Bernal-stacked,
but randomly rotated, layers. A single component D'-band is also
characteristic of the ``turbostratic''
graphite\cite{Cancado06,Pimenta07}. The differences in the Raman
scattering spectra of FLG-C-SiC and of ``turbostratic'' graphite
are however also clear. ``Turbostratic'' graphite show much wider
D'-peaks (typically 50 cm$^{-1}$ in Ref. 18) and a visible D-band,
the latter due to the appreciable disorder and/or small size of
such graphitic granulates. On the other hand, the D'-peaks in
FLG-C-SiC and ``turbostratic'' graphite are centered at very
similar energies, but characteristically higher, by about 20
cm$^{-1}$ as compared to the case of graphene. While the shift of
the D'-peak for the FLG-C-SiC (and ``turostratic'' graphite) as
compared to the case of graphene might be due to a difference in
the phonon energies characteristic of each of these systems, we
speculate it is more likely due to the possible difference in the
Fermi-velocity of Dirac cones in these systems. Recent reports
clearly show that the energy position of the D'-band, arising from
the double resonant Raman scattering process, can depend on the
Fermi-velocity and this parameter seems to be $\sim10\%$ larger
for graphene\cite{Jiang07} compared to FLG-C-SiC\cite{Sadowski06}.
While no shift is observed for the G-band, a third possible but
unlikely explanation for the observed energy shift of the D' band
could also be a natural doping of the FLG-C-SiC\cite{Yan07}.

In conclusion, we have found the majority signal of micro-Raman
scattering observed from graphitized carbon-terminated surfaces of
SiC shows a double resonant D'-band in the form of a single
component peak which indicates the single-band electronic
structure of this material. Together with previously reported
results of transport\cite{Berger06},
spectroscopic\cite{Sadowski06} and structural
investigations\cite{Hass07}, the present data confirms the
appearance of Dirac-like electronic states in FLG-C-SiC.
Remarkably, this is in spite of its multilayer character. Almost
certainly, the layers of FLG-C-SiC are electronically well
decoupled, as is the case to some extent in ``turbostratic''
graphite. This suggests that functional graphene-based devices can
be developed using the methods of thermal decomposition (epitaxy)
of silicon carbide, opening the way for many potential
applications.

We gratefully acknowledge E. Bustarret and D.K. Maude for
stimulating discussions.

\bibliography{}

\end{document}